# Neural Network Model of Pricing Health Care Insurance

**Abstract** – To pricing health insurance plan, statisticians use mathematical models to analysis customer's future health condition. General Addictive Model (GAM) is a wide-accepted method for this problem, however, it have several limitations. To solve this problem, a new method named neural network model is implemented. Compare with GAM model, neural network provide a more accurate predicting result.

## 1. Introduction

### 1.1 General information of health care insurance pricing problem

Health care insurance is a financial structure that against the risk of future medical expenses of individuals. By predicting the health care expenses of a certain group of people, an agency sets up a monthly payroll called premium. The insurance company will provide health care benefits to the individuals who pay the premium regularly. Benefits are specified by a contract between the insurance agency and its customer. Most Large companies provide health care insurance to their employee. And the government also supplies Medicare (a special health insurance) to the disables, aged people, and people who have no income. [1]

Pricing healthcare insurance is a complex procedure. It is important for a company to pricing each customer's health insurance plan individually base on his/her life condition. Several factors, including gender, age, health condition, income, job status, and the living place, would be considered [1]. The variability of price of health care insurance plan for different people is high. For people who had a severe disease, the price of health care insurance will higher than average.

Healthcare industries have several principles for pricing health insurance products [2]. Firstly, premium amount need to be adequate. The price must low enough to attract consumers while providing a reasonable amount of profit to the company itself. Secondly, premium amounts need to match the coverage that the health insurance could provide and meet the requirement of customer. Few costumers will pay for low price insurance without enough benefit. If the company charges too much money for the coverage of the health care product, it will lose its costumers. Thirdly, the premium of health insurance product need to competitive compare with the amount charged by other health insurance companies for similar products. Few costumers wanted to spend a great deal of money just for a special reasonable benefits policy. In most cases, a better price won more costumers. Finally, the health insurance of every customer must keep equality. Some customer placed much more claims than other costumers. As a result, the price for each individual is expected to be different. Thus, the health care insurance industries need to design distinct insurance plan focus on different target customer group. For each plan, health insurance companies need to construct mathematical models, predicting the future health care expenditure base on customer's health condition and other information. After consider future expenditure of customers and operation cost of company itself, statistician could set up a reasonable price and coverage of a health insurance plan. Several statistical methods were used to pricing health insurance plan [3].

### 1.2 Form History of mathematical modeling about pricing health care insurance

Many researchers have developed statistical models to estimate the future medical expenditure by analysis people's previous health status. In early 1990s, most health industries used General Linear Model (GLM) to predict future health condition for customers. GLM is a flexible extension of linear regression for a single dependent variable. It contains several explanatory variables denoted as $x_1, x_2 \cdots x_m$ and one responsive variable $g(y)$. For health insurance pricing problem, statisticians choose gender, age, working status, income, and previous insurance claims of customers as explanatory variables. And the response variable $g(y)$ could be future medical expenditure or price of health insurance of customers. Then, statisticians estimate parameters $\beta_0, \beta_1, \beta_2, \cdots \beta_m$ by a dataset of previous customers' information. The model is shown as equation (1).

$$g(y) = \beta_0 + \beta_1 x_1 + \beta_2 x_2 + \cdots + \beta_m x_m \qquad (1)$$

After constructed the GLM model, health insurance companies could predict the future medial expenditure of a new customer and price his health insurance plan [3]. Recently, GLM model is still a wide-accepted model for approximate estimate in some health insurance companies [4].

After 1995, General Additive Model (GAM) gradually entered people's vision. GAM model is developed from blending of the key characters of generalized linear model and additive model. [7] In a typical GAM model, the explanatory variables $x_1, x_2 \cdots x_m$ and parameters $\beta_0, \beta_1, \beta_2, \cdots \beta_m$ in GLM is replaced by a series of more complex functions of explanatory variables denoted as $f(x_1), f(x_2) \cdots f(x_m)$. The function $f(x)$ provides a better fitting for dataset than parameters in GLM model. It named smooth function. In the other hand, we use link function $g(E(y))$ instead of $g(y)$ as response variable of the model. The model is shown as equation (2).

$$g(E(y)) = \beta_0 + f(x_1) + f(x_2) + \cdots + f(x_m) \qquad (2)$$

To construct a GAM model, we have several steps. Firstly, assuming the data of customers is selected from a population with a specific distribution (usually normal distribution or binomial distribution). Then, we use a link function $g(E(y))$ to relating the expectation value $E(y)$ of that distribution and several explanatory variables. Secondly, we need to collect the customer's information in database of healthcare insurance company to construct the smooth function. This procedure called training. After training process, the established GAM model is well prepared for pricing the health insurance plan for a new customer. Now, GAM has become the most popular method among statistical models in the field of pricing health care insurance [5] [6].

### 1.3 Limitation of GAM model

GAM model provides a good fitting of most datasets. However, it suffers from four major limitations. Interaction of explanatory factors is the most important problem. Interaction means an effect of two or more factors to a certain result is not simply added. For example, regular smoking will increase the probability of a people to suffer from lung cancer. We denote the amount of increase as $p_1$. And drinking alcohol will also increase the probability of a person to have cancer. We denote the increase as $p_2$. Then, for a person who is always drinking and smoking, his chance to suffer from cancer is not only $p_1 + p_2$ but $p_1 + p_2 + a \times p_1 \times p_2$, where $a$ is a constant index. In other words, existence of an interaction effect implies that the effect of one explanatory variable is a function of another explanatory variable. If variables $x_1$ and $x_2$ have interaction relationship, the original GAM model will be changed:

$$g(E(y)) = \beta_0 + f(x_1) + f(x_2) + f(x_1) \times f(x_2)$$
$$+ \cdots + f(x_m) \qquad (3)$$

For a set of explanatory variables $x_1, x_2 \cdots x_m$, we need to exam the interaction effect of these variables for $\frac{m(m-1)}{2}$ times. To acquire a more accurate prediction of customer's future health condition, statisticians need to select more than 20 explanatory variables to construct the GAM model. If so, re-constructing of GAM model with interaction effect will more than 200 times. Statisticians need to exam whether each of them is significant. If we include high order interaction (the interaction effect of more than 2 variables) effect in the analysis, the problem will be even more complex. In most cases, researchers spend a lot of time on interactions of explanatory variables. They adjust the model many times to make sure it is constructed correctly. To simplify this procedure, researcher might reduce the number of explanatory variables to improve predictive ability of this model. [6]

The second disadvantage of GAM model is collinear. Collinear is a term describing two or more variables have a correlation relationship. For example, a smoking customer has a higher probability to have respiratory disease. It is unfair to charge this customer for both reasons. To solve this problem, analyst need to use factor analysis method to combine the collinearly variables into one factor, and then fitting the regression model of several factors instead of explanatory variables. It will spend a lot of time and lose the accuracy of the model. [7].

Another limitation worth mentioning is over fitting. Over fitting occurs when a statistical model is excessively complicated. A over fitting model generally has a prefect fitting for its training data but has poor predicting result. It is not always good to pursue a better fitting result. Most statistical models have a threshold. When training the model, if the fitting accuracy increase higher than this threshold, the predicting accuracy will decrease while the training accuracy increase. For the GAM model in health insurance pricing problem, analysts used to control the fitting error of predicting expense not lower than 8% to avoid over fitting. As the number of smooth parameters increase, the over fitting problem will become more obvious [4].

Besides, GAM model is restricted by parametric assumption. To acquire an outcome of customer's health condition, insurance companies need to assume that data is selected from a certain kind of distribution. However, the assumption might not exactly correct. The parametric model might suffer from the inaccurate assumption. It is difficult to estimate the parameters of model under this situation. [7]

## 2. Materials and methods

### 2.1 Neural network model

Neural network model (artificial neural network, ANN) was designed base on the structure of brain and neural system, attempting to simulate the process of human study and decision making based on a certain data set. It is a newly developed computational model in artificial intelligence field.

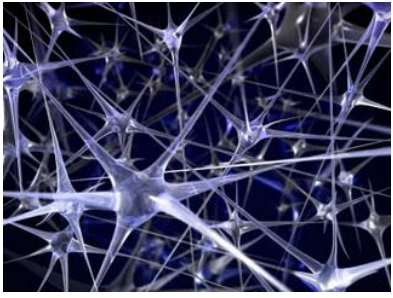

**Figure 1.** Human brain and nervous system modeled by computer system.

A neural network consists of several input variables, one or more output factors and a group of connected hidden points called inner neurons. Each input variable connects to all inner neuron in hidden layer. The value of an inner neuron is the sum of each input value multiple to a weight. The weight could be thought as the memory of human's bran; it could be changed while training. And the output value of the model denote as the sum of each inner neuron value multiple to other weight. The weights between each two points are different. The model changes its weight when training information flowing through the network. After the model was constructed, it could be used to predict the future health care expense of a new customer and determine how much insurance fee should be paid base on his current health condition [7].

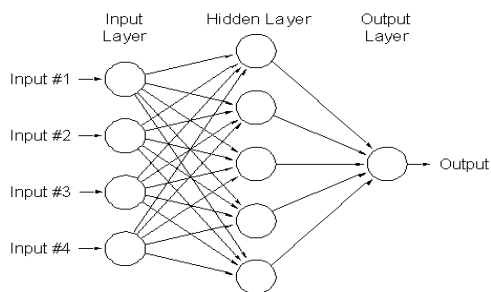

**Figure 2.** Structure plot of neural network model.

Compare with GAM model, neural network model have several advantages [16]. First of all, ANN is a nonparametric method. In other word, this model adapted sample dataset selected from population of all kinds of distributions. Therefore, statisticians could use ANN model directly without analysis the probability distribution of dataset. Secondly, because of its complex inner structure, the statisticians need not to consider interaction effect during analysis. Finally, in the training process, the changing of weight structure will automatically combine the collinear factors. In a word, by using ANN model, statisticians avoid spending time on interaction effect, collinear analysis, and assumption of population distribution. Thus, the process of constructing an artificial neural network is much simpler than GAM model and other regression model. ANN model has better performance than traditional statistical approaches on predicting of people's health condition. [9]

Neural network have two different classifications: feed forward network (FF neural network) and back propagation network (BP neural network). The structure of these two kinds of networks has slightly different. In the first step, inner points of feed forward network could only accept inputting data. And points will output the data in next step. BP neural is a more popular model in recent years. In BP neural network, each inner point could input information and output information at the same time. Neurons of BP neural network could accept feedback of next layer. This model provides a more accurate predicting result than FF neural network. Thus, we use BP network to calculate the price of health care insurance in next paragraph.[14]

## 2.2 An example of pricing healthcare insurance by neural network model

To show how neural networks works on pricing healthcare insurance problem, we run a neural network model by a set of data of 200 customer's personal information. This dataset is collected by Brockett in a research project at 2009[8].

**Table 1** the information of first 5 customers in the training data of the example. (COPD is chronic obstructive pulmonary disease).

| No. | Gender | Age | Income (per year) | Smoke | Previous Claim | Expenditure |
|---|---|---|---|---|---|---|
| 1 | female | 58 | 0 | Yes | COPD | 10,250$ |
| 2 | male | 32 | 83,000$ | No | None | 0 |
| 3 | male | 45 | 67,000$ | Yes | Lung Cancer | 148,765$ |
| 4 | female | 24 | 45,000$ | No | None | 100$ |
| 5 | female | 37 | 30,000$ | No | Diabetes | 5,200$ |

We use neural network Toolbox of Matlab software to deal with this problem. We construct a BP neural network model with 6 input points, 8 inner points (1 layer), and 1 output point. And we use half of data as training sample and other half data to exam the predicting accuracy of this model.[13]

## 3. Result and Discussion

### 3.1 Comparison of the results of neural network model and GAM model

**Table 2** the summary of accuracy of neural network model and GAM model

|  | Accuracy | Interaction analysis | Collinear analysis |
|---|---|---|---|
| ANN | 94%~107% | None | None |
| GAM | 90%~110% | Smoke | Smoke & Claim |

From table 2, we could see that ANN has a better predicting accuracy than GAM model.

**Table 3** the summary of over fitting of neural network model and GAM model

|     | Distribution assumption | Error of Over fitting threshold |
|-----|-------------------------|--------------------------------|
| ANN | None                    | 6%                             |
| GAM | Normal                  | 8%                             |

### 3.2 Limitation of neural network model

The neural network model has an excellent predicting result on pricing health care insurance problem. However, it have not solved the over fitting problem entirely. It just reduce the over fitting threshold of the model. In a GAM model, analyst needs to control the error of predicting result close to 8%. And statisticians could reduce the over fitting threshold to 6% in ANN model [9].

### 3.3 Analysis improvement and development of neural network model of healthcare insurance problem

To improve the accuracy of predicting without over fitting, analyst could try to collect more customers information as training data. The over fitting threshold will decrease while training sample is large. [10,11]